\documentclass[prd,preprint,amssymb,amsmath,showpacs]{revtex4}
\usepackage{graphicx}

\begin{document}
%
{\baselineskip0pt
\leftline{\large\baselineskip16pt\sl\vbox to0pt{\hbox{\it Department of Physics}
               \hbox{\it Osaka City  University}\vss}}
\rightline{\large\baselineskip16pt\rm\vbox to20pt{\hbox{OCU-PHYS-189}
            \hbox{AP-GR-5}
\vss}}
}
\vskip1.5cm
\title{Volume Expansion of Swiss-Cheese Universe}
%
\author{Hiroshi Kozaki}
 \email{furusaki@sci.osaka-cu.ac.jp}
\author{Ken-ichi Nakao}
 \email{knakao@sci.osaka-cu.ac.jp}
%
\affiliation{%
Department~of~Physics,~Graduate~School~of~Science,~Osaka~City~University,
Osaka 558-8585, Japan%
}
%
\begin{abstract}
In order to investigate the effect of inhomogeneities on 
the volume expansion of the universe, we study 
modified Swiss-Cheese universe model.
Since this model is an exact solution of Einstein equations,
we can get an insight into non-linear dynamics of
inhomogeneous universe from it. We find that inhomogeneities make the
volume expansion slower than that of the
background Einstein-de Sitter universe when those can be regarded as
small fluctuations in the background universe.
This result is consistent with the previous studies based on the
second order perturbation analysis.
On the other hand, if the inhomogeneities can not be treated as 
small perturbations,
the volume expansion of the universe depends on the type of
fluctuations. Although the volume expansion rate approaches 
to the background value 
asymptotically, the volume itself can be finally arbitrarily smaller
than the background one and can be larger than that of the background 
but there is an upper bound on it. 
\end{abstract}
\maketitle

\section{INTRODUCTION}       

The standard Big Bang scenario is based on an assumption of the 
homogeneous and isotropic distribution of matter and 
radiation. This assumption then leads to the Robertson-Walker 
spacetime geometry and the Friedmann-Lema\^{\i}tre (FL)
universe model through the Einstein equations. 
This model has succeeded in explaining various important observational 
facts: Hubble's expansion law, the content of light elements and 
the isotropic cosmic microwave background radiation 
(CMBR)\cite{ref:Weinberg}. 

The CMBR conversely gives a strong observational basis for the
assumption of homogeneity and isotropy of our universe by its highly
isotropic distribution together with the Copernican principle;  
we know that our universe was highly 
isotropic and homogeneous at least on the last scattering 
surface where CMBR comes from\cite{ref:Smoot}. Hence, in the early stage of 
our universe, the linear perturbation analysis in the 
FL universe model is a powerful tool 
to investigate the dynamical evolution of our universe\cite{ref:KS84}. 

In order to perform the perturbation analysis, 
we need an appropriate background universe model, 
i.e., information of the Hubble parameter, the density 
parameter and further the equation of state of the matter,  
radiation and so on, in the real universe.  
To fix the background universe, we use the observational data 
in the neighborhood of our galaxy. Especially the Hubble parameter 
is determined from the data about the distance-redshift relation 
within $100h^{-1}$Mpc except for the type 
Ia supernova\cite{{Schmidt:1998ys},{Perlmutter:1999np}}. 
However, the universe in a region within $100h^{-1}$Mpc is 
highly inhomogeneous and hence there are non-trivial prescriptions  
to identify the present inhomogeneous universe to the 
homogeneous and isotropic FL universe model. 
If those procedure are not appropriate, we might miss 
finding the correct background universe. 

It is often stated that the spatially averaged observational data 
in the vicinity of our galaxy are recognized as those of the 
background FL universe model. 
The Hubble parameter determined by the observed 
distance-redshift relation in our universe 
is regarded as the expansion rate of the volume 
of the region co-moving to matter. 
There are several researches for the effects of inhomogeneities 
on the volume expansion of the
universe\cite{ref:futamase89,ref:futamase97,ref:Tomita,ref:Russ,
ref:Mukhanov,ref:Abramo,ref:Nambu,ref:Nambu-2,ref:Nambu-3,ref:Geshnizjani}. 
Especially, Nambu applied the renormalization 
group method to the second order 
cosmological perturbation theory and claimed that the expansion of 
the dust filled universe is decelerated 
by the inhomogeneities\cite{ref:Nambu}.

In the real universe, this back reaction effect might be very 
small\cite{ref:Russ}. 
However in order to get deeper 
insight into the dynamics of the inhomogeneous universe, 
we will consider the situation in which the 
back reaction of the inhomogeneities seems to be effective. 
For this purpose, we consider the Swiss-Cheese 
universe model and investigate the volume expansion in it. 
The original Swiss-Cheese universe model is 
constructed by choosing non-overlapping spherical regions in 
the background homogeneous and isotropic dust filled 
universe and then replacing 
these regions by the Schwarzschild space-time whose mass parameter 
is identical with the ``gravitational'' 
mass of the dust fluid in the removed 
region. On the other hand, 
in this article, we consider a modified version; 
we first remove spherical regions from the homogeneous and isotropic 
dust filled universe and then fill these regions with  
spherically symmetric but inhomogeneous dust balls. 
A spherically symmetric inhomogeneous dust ball is described 
by the Lema\^{\i}tre-Tolman-Bondi (LTB) solution 
which is an exact solution of the Einstein equations, and hence  
by this procedure, we obtain an exact solution of Einstein equations, 
which represents an inhomogeneous universe. 
Using this solution, we can study non-linear effects 
of inhomogeneities on the volume expansion of the universe 
without use of perturbation analysis. 

In the LTB solution, shell crossing singularities are generic. 
Since the LTB solution is no longer valid 
after the occurrence of the shell crossing,  
we need to change the treatment if it occurs. 
As a crude approximation to describe 
the dynamics after the shell crossing, we adopt a model 
in which the shell crossing 
region is replaced by a spherical dust shell. 

This article is organized as follows. 
In section \ref{sec:SC}, we explain how to construct  
modified Swiss-Cheese universe models which are studied in this article. 
We investigate the volume expansion rate in the case of small 
perturbations in section \ref{sec:perturb} 
and in the highly inhomogeneous case in \ref{sec:non-linear}. 
In section \ref{sec:non-linear},
an alternative model is also constructed which describes the universe after
the shell crossing and investigate the dynamics of this model.
Finally, section \ref{sec:summary} is devoted to summary and 
discussion. 

We use the units in which $c=G=1$ throughout the paper.

\section{MODIFIED SWISS CHEESE UNIVERSE MODEL} \label{sec:SC}

In this section, we give a prescription to construct 
MSC universe model. 
First we consider a Einstein-de Sitter (EdS) universe and then remove 
spherical regions from it; these removed regions should not overlap 
with each other. Next these regions filled with inhomogeneous dust 
balls with the same radii and the same gravitational mass 
as those of the removed homogeneous dust balls. 
In this MSC universe model, each inhomogeneous region 
is described by the Lema\^{\i}tre-Tolman-Bondi (LTB) solution 
which is an exact solution of Einstein equations. 

LTB solution describes the dust filled spherically symmetric spacetime.
Adopting synchronous and co-moving coordinate system, 
the line element is written as
\begin{align}
 ds^{2}=&-dt^{2}+\gamma_{ij}dx^{i}dx^{j} \notag \\
       =&-dt^{2}+\frac{Y'{}^{2}(t,\chi)}{1-\chi^{2}k(\chi)}d\chi^{2}
        +Y^{2}(t,\chi)(d\theta^{2}+\sin^{2}\theta d\varphi^{2}), 
\label{eq:line-element}
\end{align}
where the prime ${}'$ denotes the differentiation 
with respect to the radial coordinate $\chi$. 
In this coordinate system, components 
of 4-velocity $u^{a}$ of a dust fluid element are 
\begin{equation}
u^{a}=(1,~0,~0,~0).
\end{equation}
The stress-energy tensor $T_{ab}$ is then given by 
\begin{equation}
 T_{ab}=\rho(t,\chi) \delta^{0}_{a} \delta^{0}_{b},
\end{equation}
where $\rho(t,\chi)$ is the rest mass density of the dust. 

Einstein equations lead to the equations for the areal radius $Y(t,\chi)$ 
and the rest mass density $\rho(t,\chi)$ of the dust;
\begin{align}
 \dot{Y}^{2} =& -\chi^{2}k(\chi) + \frac{2M(\chi)}{Y},\label{eq:einstein}  \\
 \rho =& \frac{M'(\chi)}{4\pi Y'Y^{2}},\label{eq:density}
\end{align}
where $k(\chi)$ and $M(\chi)$ are arbitrary functions 
and the dot $\dot{}$ denotes the differentiation 
with respect to $t$. 

We set $M(\chi)$ as 
\begin{equation}
M(\chi)=\frac{4\pi\rho_{0}}{3}\chi^{3}, \label{eq:mass-form}
\end{equation}
where $\rho_{0}$ is a non-negative arbitrary constant.
The above choice of $M(\chi)$ does not loose any generality. 
Eqs.~(\ref{eq:line-element})-(\ref{eq:density}) are
invariant for the rescaling of the radial coordinate $\chi$,
\begin{equation}
 \chi \rightarrow \tilde{\chi}=\tilde{\chi}(\chi).
\end{equation}
Considering this property, 
we can choose above form of $M(\chi)$ as long as $\rho Y'>0$.

\begin{figure}
 \begin{center}
  \includegraphics{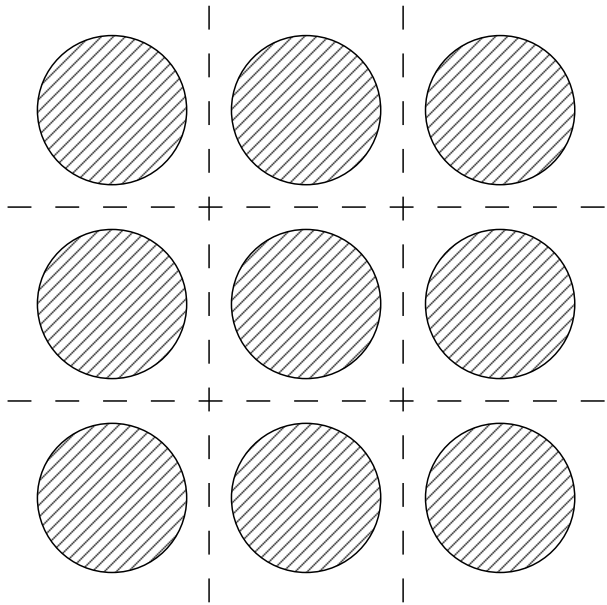}
 \end{center}
 \caption{The MSC universe model. 
 Each shaded region represents the inhomogeneity and is described by 
 LTB solution.
}
\label{fig:MSC}
\end{figure}

\begin{figure}
 \begin{center}
  \includegraphics{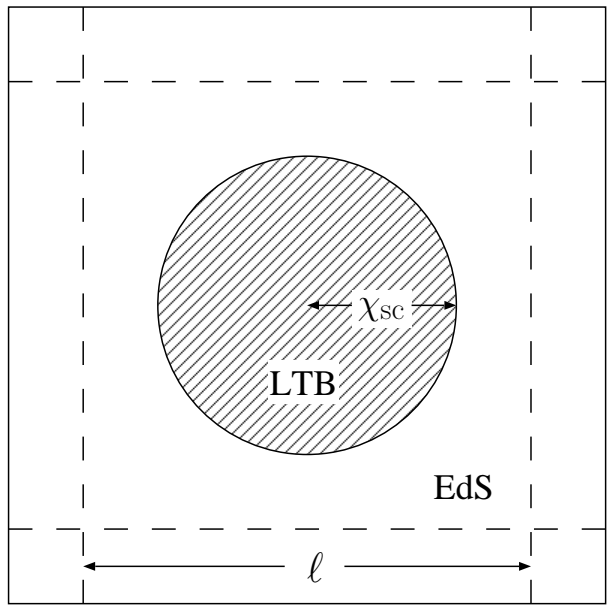}
 \end{center}
 \caption{One cubic region $\Omega$ 
 of the MSC universe model.
 $\ell$ and $\chi_{\rm sc}$ are the co-moving scales of this 
 cubic region and the inhomogeneous  region respectively.}
\label{fig:MSC-2}
\end{figure}

The solutions of eq.~(\ref{eq:einstein}) are given as follows: 

\noindent
In the region where $k(\chi)>0$,
\begin{align}
 Y=&\frac{4\pi\rho_{0}}{3k}(1-\cos\eta)\chi, \label{eq:k>0}\\
 t-t_{0}(\chi)=&\frac{4\pi\rho_{0}}{3k^{3/2}}(\eta-\sin\eta);
\label{eq:t-solution}
\end{align}
in the region where  $k(\chi)=0$,
\begin{equation}
 Y=\Bigl[ 6\pi\rho_{0} \bigl\{ t-t_{0}(\chi) \bigr\}^{2} \Bigr]^{1/3}\chi;
  \label{eq:k=0}
\end{equation}
in the region where $k(\chi)<0$,
\begin{align}
 Y=&\frac{4\pi\rho_{0}}{3|k|}(\cosh\eta-1)\chi, \label{eq:k<0}\\
 t-t_{0}(\chi)=&\frac{4\pi\rho_{0}}{3|k|^{3/2}}(\sinh\eta-\eta),
\end{align}
where $t_{0}(\chi)$ is an arbitrary function. 
Note that $t_{0}(\chi)$ is the time when a shell focusing 
singularity appears, where `shell focusing singularity' means 
$Y=0$ for $\chi>0$ and $Y'=0$ at $\chi=0$. 
In this article, we consider a region of $t>t_{0}$, 
and hence the time $t=t_{0}$ corresponds to 
the Big Bang singularity. Here, we focus on a case 
of $t_{0}=0$, i.e., simultaneous Big Bang.  
For simplicity, we consider the simplest version of MSC universe 
models shown in fig.~\ref{fig:MSC}; 
there is only one inhomogeneous spherical region at the center 
in each identical cubic region $\Omega$.
We focus on only one cubic co-moving 
region $\Omega$ (fig.~\ref{fig:MSC-2}). 
In this article, we consider a following model. 
Assuming $0<\chi_{1}<\chi_{2}<\chi_{3}<\chi_{\rm sc}$, 
\begin{equation}
 k(\chi)=
  \begin{cases}
   k_{0} & \text{for~~~} 0\leq\chi<\chi_{1}\\ 
   \dfrac{k_{0}}{2\chi^{2}}
    \left\{
     \dfrac{(\chi^{2}-\chi_{2}^{2})^{2}}{\chi_{1}^{2}-\chi_{2}^{2}}
     +\chi_{1}^{2}+\chi_{2}^{2}
   \right\} & \text{for~~~} \chi_{1}\leq\chi<\chi_{2} \rule{0pt}{20pt} \\
   \dfrac{k_{0}}{2\chi^{2}}
   \left(\chi_{1}^{2}+\chi_{2}^{2}\right)
   & \text{for~~~} \chi_{2}\leq\chi<\chi_{3} \rule{0pt}{26pt}\\
   \dfrac{k_{0}}{2\chi^{2}}\left(\chi_{1}^{2}+\chi_{2}^{2}\right) 
   \left\{
     \left(\dfrac{\chi^{2}-\chi_{3}^{2}}
      {\chi_{\rm sc}^{2}-\chi_{3}^{2}}
     \right)^{2}-1
   \right\}^{2}
   & \text{for~~~} \chi_{3}\leq\chi<\chi_{\rm sc} \rule{0pt}{26pt}
  \end{cases},
  \label{eq:region-2-3}
\end{equation}
where $k_{0}$ is constant. In order to 
guarantee $1-\chi^{2}k>0$, the following inequality should hold,
\begin{equation}
\kappa:=\dfrac{k_{0}}{2}(\chi_{1}^{2}+\chi_{2}^{2})<1.
\label{eq:kappa-def}
\end{equation}

We consider two cases; one is the case 
of $k_{0}>0$ and the other is the case of $k_{0}<0$. 
Then we investigate the volume expansion rate 
of a cubic co-moving region $\Omega$. 
The volume $V$ is defined by
\begin{equation}
V(t):=\int_{\Omega}\sqrt{\gamma}d^{3}x, \label{eq:volume-def}
\end{equation}
where $\gamma$ is the determinant of the spatial metric $\gamma_{ij}$. 
The volume expansion rate is defined as ${\dot V}/V$. 

\section{The Case of Small Fluctuations} \label{sec:perturb}

In a region where
\[
0< \frac{9t|k|^{3/2}}{2\pi\rho_{0}}\ll 1
\]
is satisfied, the areal radius 
$Y(t,\chi)$ is written in the form of power series as 
\begin{equation}
 Y(t,\chi)=a(t)\chi
        \left(
              1-\frac{1}{20}\epsilon-\frac{3}{2800}\epsilon^{2}
        \right)+O(\epsilon^{3}) , \label{eq:perturbation}
\end{equation}
where 
\begin{equation}
\epsilon(t,\chi):=\left( \frac{9t}{2\pi\rho_{0}} \right)^{2/3}k
         =\left( \frac{12t}{M(\chi)} \right)^{2/3}\chi^{2}k,
\end{equation}
and
\begin{equation}
a(t):=(6\pi\rho_{0}t^{2})^{1/3}. \label{eq:s-factor}
\end{equation}
Further, we consider the case of $|k|\chi^{2}\ll1$, which has been studied 
by Nambu by the second order perturbation analysis. 
The components of the metric tensor are written as
\begin{align}
 g_{\chi\chi}
   &= a^{2}
       \Bigg[
           1
           +\left(\dfrac{M}{12t}\right)^{2/3}\epsilon
           -\frac{1}{10}\frac{d}{d\chi}(\chi\epsilon)
          +O(\epsilon^{2})
       \Bigg], \label{eq:metric-perturbation-1}\\
 g_{\theta\theta}
   &= a^{2}\chi^{2}
       \Bigg[ 
           1-\frac{1}{10}\epsilon+O(\epsilon^{2}) 
       \Bigg], \label{eq:metric-perturbation-2}\\
 g_{\varphi\varphi}
   &= a^{2}\chi^{2}\sin^{2}\theta
      \Bigg[
         1 -\frac{1}{10}\epsilon+O(\epsilon^{2})
      \Bigg]. \label{eq:metric-perturbation-3}
\end{align}
From the above equations, it is easily seen that 
in the limit $\epsilon\rightarrow 0$ with $t$ fixed, 
the metric tensor becomes that of EdS. Since  
the outside region is EdS universe, 
$\epsilon$ should vanish at the boundary $\chi=\chi_{\rm sc}$  
by the continuity of the metric tensor. 

To compare our result with the study by Nambu\cite{ref:Nambu}, 
we impose conditions that spatial averages of 
the Cartesian components of the metric tensor and of the density agree 
with those of the \textit{background} EdS universe  up to the first
order of $\epsilon$. 

The spatial average of a quantity $F$ is defined as follows:
\begin{equation}
\langle F \rangle := \left(\int_{\Omega}d^{3}x\right)^{-1}
\int_{\Omega} F d^{3}x. 
\end{equation}

We consider a Cartesian coordinate system $(x,y,z)$ which 
is related to the spherical polar coordinate system 
$(\chi,\theta,\varphi)$ in the ordinary manner as 
\begin{align}
 x&=\chi\sin\theta\cos\varphi, \notag \\
 y&=\chi\sin\theta\sin\varphi, \notag \\
 z&=\chi\cos\theta. \nonumber
\end{align}
Then the spatial averages of Cartesian components of the spatial 
metric $\gamma_{ij}$ are obtained as
\begin{align}
\langle \gamma_{ij} \rangle
&=a^{2}\left[
         1+\dfrac{4\pi}{3\ell^{3}}
         \int_{0}^{\chi_{\rm sc}}
           \left\{
              \left( \dfrac{M}{12t} \right)^{2/3}\epsilon\chi^{2}
             -\dfrac{1}{10}\dfrac{d}{d\chi}(\chi^{3}\epsilon)
           \right\}
         d\chi
       \right]\delta_{ij}
  +O(\epsilon^{2}) \notag \\
&=(6\pi\rho_{0}t^{2})^{2/3}
  \left( 
    1+\dfrac{4\pi}{3\ell^{3}}\int_{0}^{\chi_{\rm sc}}k\chi^{4}d\chi
  \right)\delta_{ij}+O(\epsilon^{2}),
\end{align}
where we have used $\epsilon=0$ at $\chi=\chi_{\rm sc}$. 
Therefore we should define the scale factor $a_{\textsc{b}}(t)$ 
of the background EdS universe as
\begin{equation} 
a_{\textsc{b}}(t)
:=a(t)\left(
  1+\dfrac{4\pi}{3\ell^{3}}\int_{0}^{\chi_{\rm sc}}k\chi^{4}d\chi
      \right)^{1/2}, \label{eq:linear-s-factor}
\end{equation}
The above equation means that although the outside homogeneous region 
has the same geometry of EdS universe, it does not agree with 
{\it the background} EdS universe as long as
\begin{equation}
 \int_{0}^{\chi_{\mbox{\scriptsize{sc}}}}k\chi^{4}d\chi \neq 0. 
                                                 \label{eq:condition-k}
\end{equation}

The rest mass density $\rho$ of the dust is written 
in the form of a power series with respect to $\epsilon$ as
\begin{equation}
\rho=\dfrac{1}{6\pi t^{2}}\left\{1+\dfrac{1}{20\chi^{2}}
\dfrac{d}{d\chi}(\chi^{3}\epsilon)\right\}+O(\epsilon^{2}).
\end{equation}
The spatial average of $\rho$ is obtained as
\begin{equation}
\langle\rho\rangle =\dfrac{1}{6\pi t^{2}}+O(\epsilon^{2}).
\end{equation}
Hence the background energy density $\rho_{\textsc{b}}$ is defined by 
\begin{equation}
\rho_{\textsc{b}}(t):=\dfrac{1}{6\pi t^{2}}. \label{eq:linear-density}
\end{equation}
Here note that the background Hubble equation
\begin{equation}
\left(\dfrac{{\dot a}_{\textsc{b}}}{a_{\textsc{b}}}\right)^{2}
=\dfrac{8\pi}{3}\rho_{\textsc{b}}, \label{eq:Hubble-eq}
\end{equation}
holds.

By eq.~(\ref{eq:perturbation}), the 3-dimensional volume element
$\sqrt{\gamma}$ is written as
\begin{multline}
 \sqrt{\gamma}
 = a^{3}\chi^{2}\sin\theta
     \biggl[
        1
       -\frac{1}{20\chi^{2}}\frac{d}{d\chi}(\chi^{3}\epsilon)
        +\frac{1}{2}\chi^{2}k(\chi)
         +\frac{1}{700\chi^{2}}\frac{d}{d\chi}(\chi^{3}\epsilon^{2}) \\
  -\frac{1}{40}k(\chi)\frac{d}{d\chi}(\chi^{3}\epsilon)
        +\frac{3}{8}\chi^{4}k^{2}(\chi)
     \biggr]+O(\epsilon^{3}).
\end{multline}
Using the above equation, the volume $V$ defined by 
eq.~(\ref{eq:volume-def}) is obtained as 
\begin{equation}
 V(t)= a_{\textsc{b}}^{3}(t)
       \left(
         \ell^{3}+V_{1}+V_{2}t^{2/3}
       \right)+O(\epsilon^{3}), \label{eq:linear-volume}
\end{equation}
where
\begin{align}
 V_{1}&:= \dfrac{3\pi}{2}
           \int_{0}^{\chi_{\rm sc}}
             \chi^{6}k^{2}
           d\chi 
         -\dfrac{2\pi^{2}}{3\ell^{6}}
           \left(
             \int_{0}^{\chi_{\rm sc}}
               k\chi^{4}
             d\chi
           \right)^{2}, \\
 V_{2}&:=-\dfrac{\pi}{20}
           \left( \frac{9}{2\pi\rho_{0}} \right)^{2/3}
            \int_{0}^{\chi_{\rm sc}} k^{2}\chi^{4} d\chi <0.
\end{align}
In eq.~(\ref{eq:linear-volume}), $a_{\textsc{b}}^{3}\ell^{3}$
is the 3-dimensional volume measured by background EdS 
geometry and extra terms come form inhomogeneities.
These terms do not include first order perturbations, 
but come from second order perturbations. 

The volume expansion rate is given by
\begin{equation}
   \frac{\dot{V}}{V}
 = 3\frac{\dot{a}_{\textsc{b}}}{a_{\textsc{b}}}
  +\frac{2V_{2}}{3\ell^{3}}t^{-1/3}+O(\epsilon^{3}).
\end{equation}
The first term of the R.H.S. in the above equation 
corresponds to the background part. 
On the other hand, the second term implies 
that the back-reaction of inhomogeneities decelerates 
the volume expansion.
It is worthwhile to note that this result does not depend on 
the detailed functional form of $k(\chi)$. 

\section{The Case of Non-Linear Fluctuations}\label{sec:non-linear}

In this section, we study the cases where $|k|\chi^{2}$ is 
not necessarily much smaller than unity. 
As in the case treated in the previous section, in 
order to specify an effect due to inhomogeneity, 
we need a background homogeneous 
cubic region to be compared with an inhomogeneous one. 
However the background homogeneous universe 
introduced in the previous section 
is not appropriate for the non-linear case; for example, 
too small $k$ makes the background scale factor $a_{\textsc{b}}$ 
defined in eq.~(\ref{eq:linear-s-factor}) negative. 
In order to introduce an appropriate background, 
we consider the rest mass $M_{\rm R}$ defined by 
\begin{align}
 M_{\rm R}
  &:= \int_{\Omega} \rho u^{0}\sqrt{-g} d^{3}x \notag \\
  &= 4\pi
      \int_{0}^{\chi_{\rm sc}} 
           \dfrac{\rho Y'Y^{2}}{\sqrt{1-\chi^{2}k}}
      d\chi
    +\rho_{0}
      \left(
        \ell^{3}-\dfrac{4\pi}{3}\chi_{\rm sc}^{3}
      \right) \notag \\
  &= \rho_{0}\ell^{3}
      \left\{
         1+\dfrac{4\pi}{\ell^{3}}
          \int_{0}^{\chi_{\rm sc}}
            \left(\dfrac{1}{\sqrt{1-\chi^{2}k}}-1\right)\chi^{2}
          d\chi
      \right\}, \label{eq:M-def}
\end{align}
where $g$ is the determinant of the metric tensor of spacetime. 
$M_{\rm R}$ is a conserved quantity by virtue of the continuity 
of the rest mass density, $\partial_{a}(\rho u^{a}\sqrt{-g})=0$, where 
$\partial_{a}$ is a partial derivative. 
We introduce a background as a cubic region with the same rest 
mass as the corresponding inhomogeneous cubic region. 
Note that in general, the rest mass of the dust $M_{\rm R}$ in a cubic 
co-moving region disagrees with that of the original EdS universe. 
Hence a cubic region of the original EdS universe is not background. 

The rest mass density $\rho_{\textsc{b}}$ and 
scale factor $a_{\textsc{b}}$ of the background 
are introduced in the following manner,
\begin{equation}
\rho_{\textsc{b}}a_{\textsc{b}}^{3}\ell^{3}:=M_{\rm R}. \label{eq:relation}
\end{equation}
Even if we specify $M_{\rm R}$, $\rho_{\textsc{b}}$ and $a_{\textsc{b}}$ 
are not fixed completely; we need one more condition. 
Here we impose a condition in which the volume $V(t)$ 
approaches to the volume 
$a_{\textsc{b}}^{3}(t)\ell^{3}$ of the background 
cubic region for $t\rightarrow0$. 
In the limit of $t\rightarrow0$, the volume $V$ behaves as
\begin{align}
V&= a^{3}\ell^{3}
   +4\pi
    \int_{0}^{\chi_{\rm sc}}
      \left(
        \dfrac{Y'Y^{2}}{\sqrt{1-\chi^{2}k}}-a^{3}\chi^{2}
      \right)
    d\chi \notag \\
 &\longrightarrow 
    a^{3}\ell^{3}
     \left\{
       1
      +\dfrac{4\pi}{\ell^{3}} 
        \int_{0}^{\chi_{\rm sc}}
          \left( \dfrac{1}{\sqrt{1-\chi^{2}k}}-1 \right)\chi^{2}
        d\chi
     \right\}
\end{align}
Hence the background scale factor $a_{\textsc{b}}$ is given by
\begin{equation}
a_{\textsc{b}}(t):=a\left\{1+\dfrac{4\pi}{\ell^{3}}
\int_{0}^{\chi_{\rm sc}}\left(\dfrac{1}{\sqrt{1-\chi^{2}k}}-1\right)
\chi^{2}d\chi\right\}^{1/3}. \label{eq:a-def}
\end{equation}
Here note that in the case of $\chi^{2}|k|\ll1$, 
the above definition of $a_{\textsc{b}}$ agrees with 
eq.~(\ref{eq:linear-s-factor}) up to the first order of $\chi^{2}k$. 
From eqs.~(\ref{eq:M-def}), (\ref{eq:relation}) and (\ref{eq:a-def}), 
we find that the background rest mass density $\rho_{\textsc{b}}$ is 
completely the same as eq.~(\ref{eq:linear-density}). 
We can easily see that the background Hubble equation 
(\ref{eq:Hubble-eq}) also holds.

From eq.~(\ref{eq:density}), we can easily see that if $Y'$ vanishes, 
the rest mass density $\rho$ becomes infinite and hence a singularity
forms there. This is called shell crossing singularity. 
Hellaby and Lake showed that 
a necessary and sufficient condition for the appearance of a shell 
crossing singularity is \cite{ref:Hellaby-Lake} 
\begin{align}
k'&> 0~~~~~{\rm for~the~region}~~k>0, \\ 
\left( \chi^{2}k \right)'&> 0~~~~~{\rm for~the~region}~~k\leq0. 
\label{eq:condition}
\end{align}
In the case of $k_{0}>0$, the first condition does not hold 
and hence a shell crossing singularity does not appear. On the other 
hand, in the case of $k_{0}<0$, 
shell crossing singularities $Y'=0$ necessarily 
appear since $(\chi^{2}k)'>0$ in the  region 
$\chi_{3}\leq\chi<\chi_{\rm sc}$. 

\subsection{The Case of $k_{0}>0$}

\begin{figure}[ht]
 \begin{center}
  \includegraphics[keepaspectratio]{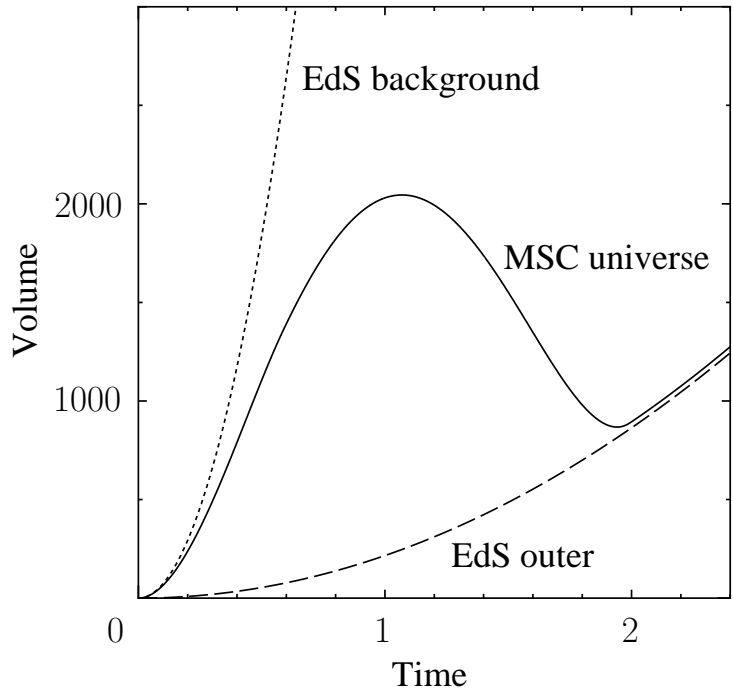}
  \caption{ \label{fig:positive}%
  Volume expansion with positive $k(\chi)$. 
  The dotted line is the temporal evolution of the co-moving volume
  measured by the background EdS geometry 
  and the dashed line is that measured by the outer EdS geometry.
  The time is set to unity, 
  when the scale of the spherical inhomogeneous region $a(t)\chi_{\rm sc}$
  agrees with that of the horizon scale $a(t)/\dot{a}(t)$.
  The scale factor at $t=1$ is also set to unity 
  and the co-moving scale $\ell$ is set to $4\chi_{\rm sc}$.
  Hence the co-moving volume measured by the outer EdS geometry at $t=1$ 
  is $\ell^3(=6^3=216)$.}
 \end{center}
\end{figure}

To estimate the volume $V$, we rewrite it in the form,  
\begin{equation}
V=a^{3}\left(\ell^{3}-\dfrac{4\pi}{3}\chi_{\rm sc}^{3}\right)
+4\pi\int_{0}^{Y_{\rm sc}}\dfrac{Y^{2}dY}{\sqrt{1-\chi^{2}k}},
\label{eq:volume-2}
\end{equation}
where $Y_{\rm sc}:=a(t)\chi_{\rm sc}$. 
From eq.~(\ref{eq:k>0}), we can see that $Y$ vanishes
at $\eta=2\pi$ by gravitational collapse, and hence by
substituting $\eta=2\pi$ into eq.~(\ref{eq:t-solution}), the singularity
formation time $t=t_{\rm sg}(\chi)$ is obtained as
\begin{equation}
t=t_{\rm sg}(\chi)=\dfrac{8\pi^{2}\rho_{0}}{3k^{3/2}}.
\end{equation}
Denoting the inverse function of $t_{\rm sg}(\chi)$ by 
$\chi_{\rm sg}(t)$, the region of 
$0\leq \chi \leq\chi_{\rm sg}(t)<\chi_{\rm sc}$ 
has already collapsed at time $t$ larger than 
$8\pi^{2}\rho_{0}/3k_{0}^{3/2}$. 
$\chi_{\rm sg}(t)$ approaches to $\chi_{\rm sc}$ for $t\rightarrow
\infty$ asymptotically. Here note that in the 
integrand of eq.~(\ref{eq:volume-2}), 
$\chi=\chi_{\rm sg}(t)$ at $Y=0$ 
and $\chi=\chi_{\rm sc}$ at $Y=Y_{\rm sc}$. Since $\chi^{2}k(\chi)$ is
decreasing function with respect to $\chi$ in the region 
$\chi_{3}<\chi<\chi_{\rm sc}$ 
and vanishes just at $\chi=\chi_{\rm sc}$,  we find
that $0 < \chi^{2}k(\chi)\leq \chi_{\rm sg}^{2}k(\chi_{\rm sg})$ holds
in the integrand at sufficiently large $t$. 
This means that $\chi^{2}k(\chi)$ also approaches to zero asymptotically, 
since $\chi_{\rm sg}^{2}k(\chi_{\rm sg})\rightarrow
\chi_{\rm sc}^{2}k(\chi_{\rm sc})=0$ for $t\rightarrow\infty$, and thus 
\begin{equation}
V\longrightarrow a^{3}\left(\ell^{3}-\dfrac{4\pi}{3}\chi_{\rm sc}^{3}\right)
+4\pi\int_{0}^{Y_{\rm sc}}Y^{2}dY=a^{3}\ell^{3}. \label{eq:volume-3}
\end{equation}
This equation means that 
the volume expansion rate approaches to the background value
asymptotically, i.e.,
\begin{equation}
\dfrac{\dot{V}}{V}\longrightarrow
 3\dfrac{\dot{a}}{a}=3\dfrac{\dot{a}_{\textsc{b}}}{a_{\textsc{b}}}
\text{~~~for~} t\rightarrow\infty.
\end{equation}
However the volume itself may approach to much different value from
the background one $a_{\textsc{b}}^{3}\ell^{3}$. 
$\sqrt{1-\chi^{2}k}$ can be made arbitrarily smaller than unity
in the region $\chi_{2}\leq\chi<\chi_{3}$; in the limit of 
$\kappa\rightarrow1$, $\sqrt{1-\chi^{2}k}\rightarrow 0$ in this region 
(see eqs.~(\ref{eq:region-2-3}) and 
(\ref{eq:kappa-def})).
Thus, if we set $\kappa$ to be very close to unity, 
we obtain 
\begin{equation}
a_{\textsc{b}} \sim a\left(\dfrac{4\pi}{\ell^{3}}
\int_{0}^{\chi_{\rm sc}}
\dfrac{\chi^{2}d\chi}{\sqrt{1-\chi^{2}k}}\right)^{1/3}
\gg a.
\end{equation}
In this case, the volume $V$ approaches to the value 
much smaller than 
the background one, asymptotically.
The volume expansion is also much different
from that of the background in the intermediate stage 
(see fig.~\ref{fig:positive}).
\subsection{The Case of $k_{0}<0$}

\begin{figure}[ht]
 \begin{center}
  \includegraphics[keepaspectratio]{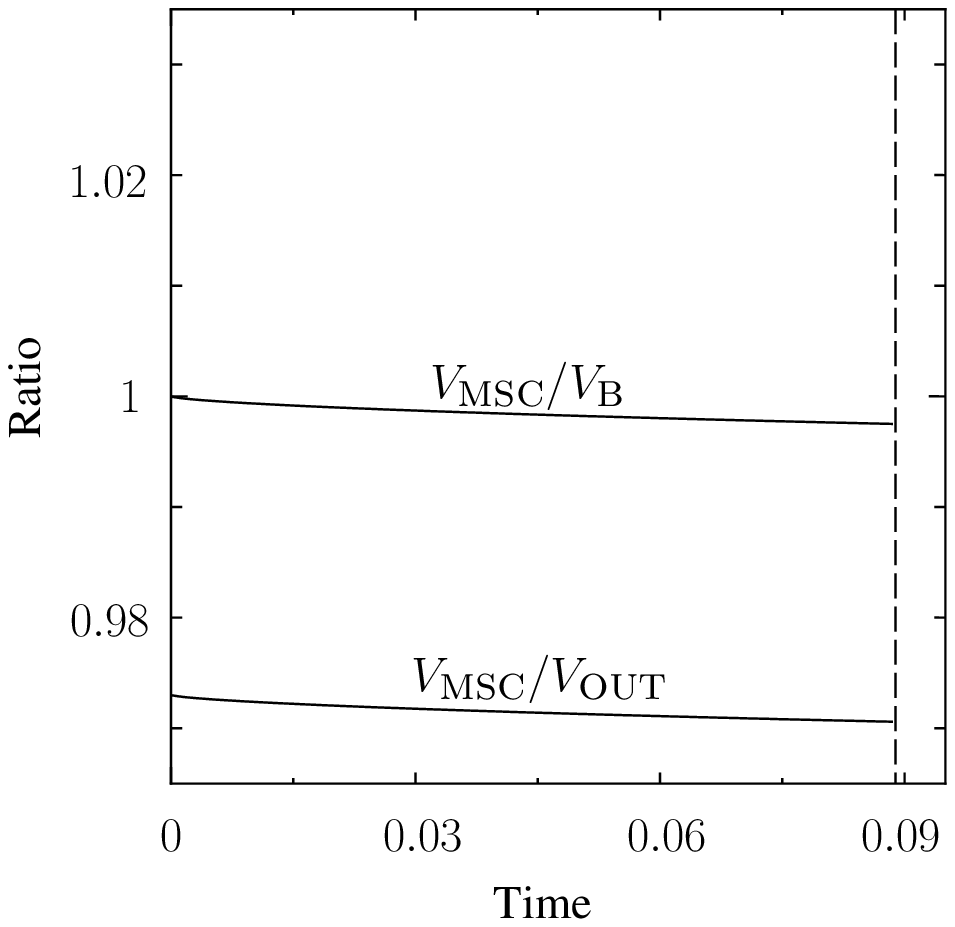}
  \caption{\label{fig:ratio.before}%
  Temporal evolution of the ratios before shell crossing
  singularity appears. 
  The time is set to unity, 
  when the scale of the spherical inhomogeneous region $a(t)\chi_{\rm sc}$
  agrees with that of the horizon scale $a(t)/\dot{a}(t)$.
  Dashed line corresponds to the shell crossing time.
  $V_{\textsc{msc}}$ is the volume of the co-moving cubic region
  $\Omega$ in modified Swiss Cheese universe.
  $V_{\textsc{b}}$ is the volume of $\Omega$ measured by the background
  EdS geometry, i.e., $V_{\textsc{b}}=a_{\textsc{b}}^{3}\ell^{3}$.
  $V_{\textsc{out}}$ is the volume of $\Omega$ measured by the outer
  EdS geometry, i.e., $V_{\textsc{out}}=a^{3}\ell^{3}$.}
 \end{center}
\end{figure}

As mentioned in the above, shell 
crossing singularities $Y'=0$ necessarily appear in this model. 
Before it, the volume expansion is shown in fig.~\ref{fig:ratio.before}.
For $k(\chi)<0$, background scale factor $a_{\textsc{b}}$ cannot differ
from original one  very much.
We plot the ratio of the volume $V(t)$ to 
$a^{3}\ell^{3}$ and $a_{\textsc{b}}^{3}\ell^{3}$.
We find that the inhomogeneities decelerate the volume expansion 
before shell crossing singularity appears.

Here, we investigate the volume expansion rate 
after the appearance of this shell crossing singularity.
The structure formed by the shell crossing 
depends on what is approximated by the dust matter. 
In case that the dust matter is extremely cold fluid, 
a shock wave will form after the shell crossing.
If the dust matter consists of collisionless particles,
a spherical wall will form. When the width of the shock wave or 
of the wall is much smaller than the radius of it, we can treat the shock 
front or the wall as a timelike singular hypersurface, 
where `timelike' means that the unit normal vector $n^{a}$ 
to the hypersurface is spacelike, i.e., $n^{a}n_{a}=1$. 

A timelike singular hypersurface is characterized 
by its surface-stress-energy tensor defined by
\begin{equation}
S_{ab}:=\lim_{\varepsilon\rightarrow0}\int_{-\varepsilon}^{\varepsilon}
T_{cd}h_{a}^{c}h_{b}^{d}dx
\end{equation}
where $x$ is a Gaussian coordinate ($x=0$ on the hypersurface) in the 
direction of the normal vector $n^{a}$, 
and $h_{a}^{c}:=\delta_{a}^{c}-n_{a}n^{c}$ is a projection operator. 
A timelike singular hypersurface with a surface-stress-energy tensor 
of the form
\begin{equation}
S_{ab}=\sigma v_{a}v_{b},
\end{equation}
is called a world sheet generated by a trajectory of a dust shell, 
where $\sigma$ is the surface-energy density and $v_{a}$ is 
the 4-velocity of an infinitesimal surface element of the 
dust shell. Hereafter we focus on this case. 

\begin{figure}
 \begin{center}
   \includegraphics{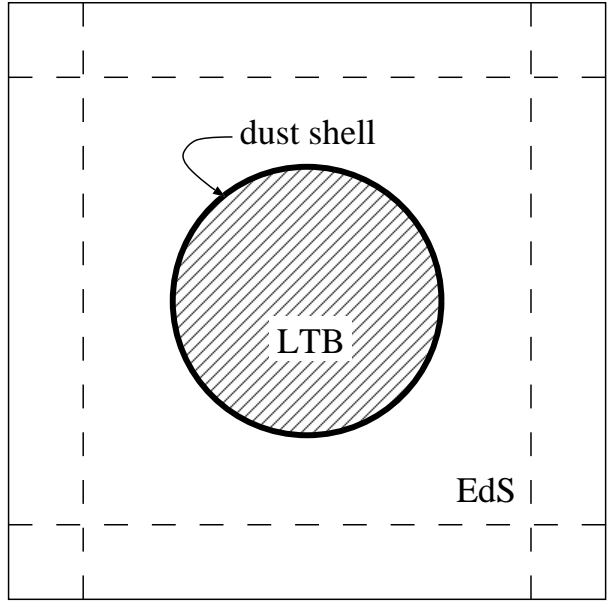}
  \caption{\label{fig:configuration} \\
  The schematic configuration of the shell. }
 \end{center}
\end{figure}

In order to get an insight into the dynamics after shell 
crossing, we study the volume expansion of a cubic region 
with a spherically symmetric dust shell. 
In the case of $k_{0}<0$,  $k(\chi)$ is negative in the 
LTB region; $(\chi^{2}k)'\leq0$ in the inner region $0\leq\chi\leq\chi_{3}$, 
while $(\chi^{2}k)'>0$ in the outer region 
$\chi_{3}<\chi<\chi_{\rm sc}$. 
In accordance with eq.~(\ref{eq:condition}), shell crossing 
necessarily occur in the outer region and we assume that 
this region collapses into a dust shell. Hence we focus on a situation 
in which $(\chi^{2}k)'<0$ holds inside the dust shell. 
The model is constructed by enclosing the interior LTB 
region by a spherically symmetric timelike singular 
hypersurface (see fig.~\ref{fig:configuration}). 

The analysis by using the dust shell model works only before 
the dust shell reaches the boundary of the cubic co-moving 
region $\Omega$. When the dust shell reaches the boundary, it collides with 
other dust shells centered in surrounding cubic regions. 
In this article, we do not consider the dynamics 
after collisions of the dust shells and assume that 
there is enough time before collisions of dust shells. 

The interior region of the dust shell is described by the LTB solution 
with a line element, 
\begin{equation}
 ds_{-}^{2}
 = -dt_{-}^2
    +\dfrac{{Y'}^{2}(t_{-},\chi_{-})}{1-\chi_{-}^{2}k(\chi_{-})}d\chi_{-}^{2}
    +Y^{2}(t_{-},\chi_{-})
     \left( 
       d\theta^{2}+\sin^{2}\theta d\varphi^{2}
     \right),
\end{equation}
where the prime ${}'$ means the derivative with respect to $\chi_{-}$.
The equation for the areal radius $Y$ is given 
by the same equation as eq.~(\ref{eq:einstein}), i.e., 
\begin{equation}
 \dot{Y}^{2}=-\chi_{-}^{2}k(\chi_{-})+\dfrac{2M(\chi_{-})}{Y},
\label{eq:density-LTB}
\end{equation}
where the dot $\dot{}$ means the derivative with respect to $t_{-}$. 
Here we focus on the late time behavior of the volume expansion. 
In this case, since $Y$ is monotonically increasing with respect to 
$t_{-}$, the ``gravitational potential'' $2M(\chi_{-})/Y$ 
becomes much smaller than $-\chi_{-}^{2}k(\chi_{-})$ 
and hence we ignore this potential term in eq.~(\ref{eq:density-LTB}). 
This approximation corresponds to an assumption in which the interior region
of the dust shell is described by Minkowski geometry. 
Then the solution for $Y$ is easily obtained as 
\begin{equation}
Y(t_{-},\chi_{-})=\left\{ -\chi_{-}^{2}k(\chi_{-}) \right\}^{1/2}t_{-}.
\label{eq:Y-solution} 
\end{equation}

The line element of the outer region is 
written as 
\begin{equation}
 ds_{+}^{2}
 = -dt_{+}^{2}
    +a^{2}(t_{+})
      \left\{
         d\chi_{+}^{2}
        +\chi_{+}^{2}
          \left(
            d\theta^{2}+\sin^{2}\theta d\varphi^{2}
          \right)
      \right\}, \label{eq:FRW}
\end{equation}
and Einstein equations reduce to the Hubble equation as
\begin{equation}
 (\dot{a}\chi_{+})^{2}=\frac{M_{+}(\chi_{+})}{a\chi_{+}}, 
                                  \label{eq:density-FRW}
\end{equation}
where $M_{+}(\chi_{+})$ is related to the rest mass density $\rho_{+}$ 
of the outer region as
\begin{equation}
 M_{+}(\chi_{+})
 = \frac{4\pi}{3}\rho_{+}a^{3}\chi^{3}_{+}.
\label{eq:mass}
\end{equation}
Note that by virtue of the spherical symmetry, angular coordinates, 
$\theta$ and $\varphi$, are common for both interior and exterior 
regions of the dust shell. 

A physically and geometrically clear prescription to treat a 
timelike singular 
hypersurface has been presented by Israel\cite{ref:Israel}. 
In his prescription, junction conditions on metric tensor across 
the singular hypersurface lead to equations 
to determine the singular hypersurface itself, i.e., 
the equation of motion of a dust shell in our case. 
Using his prescription, dynamics of a vacuum void 
surrounded by a spherical dust shell in expanding universe has been 
analyzed by Maeda and Sato\cite{ref:Maeda}. We can use their results
since the situation considered here is completely the same as theirs.

By virtue of the spherical symmetry of the system considered here, 
the trajectory of the dust shell is given by
\begin{equation}
t_{-}=t_{\rm s-}(t_{+}),~~~ 
\chi_{\pm}=\chi_{\rm s\pm}(t_{+}),~~~\theta={\rm constant}~~~{\rm and}~~~
\varphi={\rm constant},
\end{equation}
where the time coordinate $t_{+}$ in the exterior region 
has been adopted as an independent temporal variable. 
The areal radius $R$ of the dust shell is then given by
\begin{equation}
R(t_{+}):=a\left(t_{+}\right)
\chi_{\rm s+}=Y_{-}\left(t_{\rm s-},\chi_{\rm s-}\right)
\end{equation}

\begin{figure}[ht]
 \begin{center}
  \includegraphics{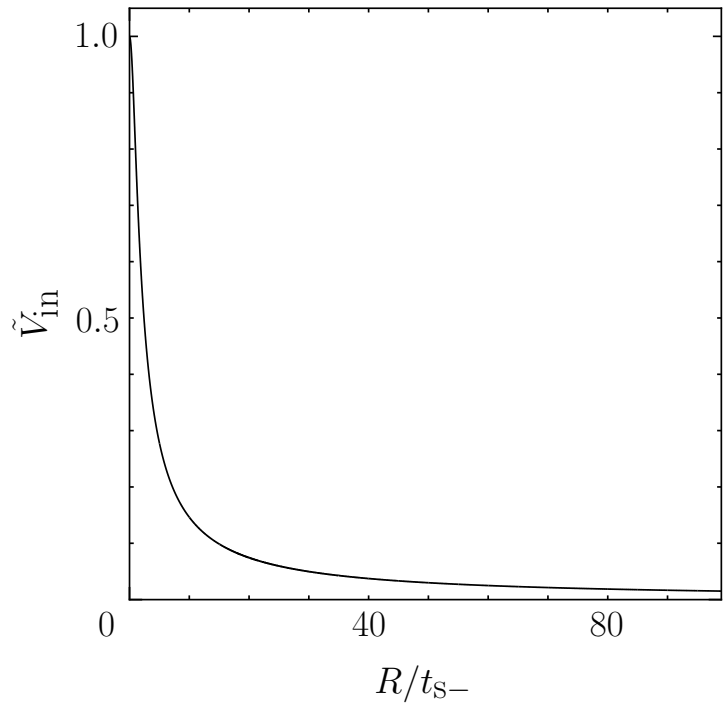}
  \caption{\label{fig:V_in}Behavior of $\tilde{V}_{\textrm{in}}$ as
  a function of $R/t_{\textrm{s}-}$. }
 \end{center}
\end{figure}

Maeda and Sato derived a differential 
equation for the areal radius $R$ of the dust shell as\cite{ref:Maeda}
\begin{equation}
\dfrac{d^{2}R}{dt_{+}^{2}}
  = \frac{1}{2R}
     \left\{-\left(1+VV_{H}+2V^{2}+V_{H}^{2}\right)
     +(1-4V^{2})(1+2VV_{H}+V_{H}^{2})^{1/2}\right\},\label{eq:R-EOM}
\end{equation}
where 
\begin{align}
 H&:= \dfrac{\dot{a}(t_{+})}{a(t_{+})}, \\
 V_{H}&:= HR, \\
 V&:=\dfrac{dR}{dt_{+}}-V_{H}.
\end{align}
Using a solution of eq.~(\ref{eq:R-EOM}), the 
coordinate radius $\chi_{\rm s+}$ of the dust shell 
in the outer region is given by
\begin{equation}
\chi_{\rm s+}=\dfrac{R}{a(t_{+})}.
\end{equation}
Equations for $t_{\rm s-}$ and $\chi_{\rm s-}$ on the dust shell are 
given by 
\begin{align}
 \dfrac{dt_{\rm s-}}{dt_{+}}
 = &\left\{1-\chi_{\rm s-}^{2}k(\chi_{\rm s-})\right\}^{1/2}
     \left\{
        1-a^{2}(t_{+})\left(\dfrac{d\chi_{\rm s+}}{dt_{+}}\right)^{2}
       +\left(\dfrac{dR}{dt_{+}}\right)^{2}
     \right\}^{1/2} \notag \\
   &-\left\{-\chi_{\rm s-}^{2}k(\chi_{\rm s-})\right\}^{1/2}
    \dfrac{dR}{dt_{+}}, \label{eq:t-EOM}\\ 
 \dfrac{d\chi_{\rm s-}}{dt_{+}}
 = &\dfrac{1}{Y_{-}'(t_{\rm s-},\chi_{\rm s-})}
     \left(
        \dfrac{dR}{dt_{+}}
       -{\dot Y}_{-}(t_{\rm s-},\chi_{\rm s-})\dfrac{dt_{\rm s-}}{dt_{+}}
     \right),
\end{align}
where the dot $\dot{ }$ and the prime $'$ are derivatives 
with respect to $t_{-}$ and $\chi_{-}$, respectively. 
The volume $V_{\rm in}$ inside the dust shell is written as
\begin{align}
 V_{\rm in}
 &= 4\pi
     \int_{0}^{\chi_{\rm s-}}
       \dfrac{Y'Y^{2}}{\sqrt{1-\chi^{2}k(\chi)}}
     d\chi
   =4\pi
     \int^{R}_{0}
       \dfrac{Y^{2}}{\sqrt{1+Y^{2}/t_{\rm s-}^{2}}}
     dY \notag \\
 &= 2\pi t_{\rm s-}^{3}
     \left[
        \dfrac{R}{t_{\rm s-}}\sqrt{1+\left(\dfrac{R}{t_{\rm s-}}\right)^{2}}
       -\ln
         \left\{
            \dfrac{R}{t_{\rm s-}}
           +\sqrt{ 1+\left( \dfrac{R}{t_{\rm s-}} \right)^{2} }
         \right\}
     \right], \label{eq:V-in}
\end{align}
where we have used eq.~(\ref{eq:Y-solution}) in the second equality 
to estimate $\chi^{2}k(\chi)$ in the integrand. 
We consider a normalized volume ${\tilde V}_{\rm in}$ by 
the volume of the removed homogeneous dust ball 
in the original EdS universe, 
\begin{equation}
{\tilde V}_{\rm in}:=\dfrac{V_{\rm in}}{4\pi R^{3}/3}=
\dfrac{3}{2}\left(\dfrac{t_{\rm s-}}{R}\right)^{3}\left[\dfrac{R}{t_{\rm s-}}
\sqrt{1+\left(\dfrac{R}{t_{\rm s-}}\right)^{2}}
-\ln\left\{\dfrac{R}{t_{\rm s-}}+\sqrt{1+\left(\dfrac{R}{t_{\rm s-}}\right)^{2}}
\right\}\right]. \label{eq:V-tilde}
\end{equation}
${\tilde V}_{\rm in}$ is 
monotonically decreasing function of $R/t_{\rm s-}$; 
it approaches to unity 
in the limit of $R/t_{\rm s-}\rightarrow0$, while 
it vanishes in the limit $R/t_{\rm s-}\rightarrow\infty$ 
(see fig.~\ref{fig:V_in}). 
In order to see temporal behavior of 
${\tilde V}_{\rm in}$, we need to solve eqs.~(\ref{eq:R-EOM}) and 
(\ref{eq:t-EOM}). 

\begin{figure}[ht]
 \begin{center}
  \includegraphics[keepaspectratio]{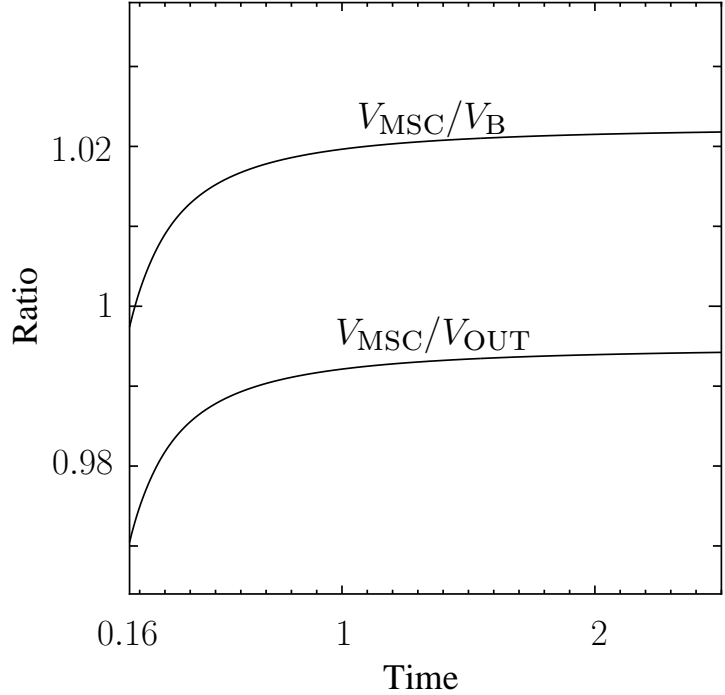}
  \caption{ \label{fig:ratio.after}%
  Temporal evolution of the ratios after shell crossing
  singularity appears. %
  The time is set to unity, 
  when the scale of spherical inhomogeneous region $a(t)\chi_{\rm sc}$
  agrees with that of the horizon scale $a(t)/\dot{a}(t)$.
  $V_{\textsc{msc}}$ is the volume of the co-moving cubic region
  $\Omega$ in modified Swiss Cheese universe.
  $V_{\textsc{b}}$ is the volume of $\Omega$ measured by the background
  EdS geometry, i.e., $V_{\textsc{b}}=a_{\textsc{b}}^{3}\ell^{3}$.
  $V_{\textsc{out}}$ is the volume of $\Omega$ measured by the outer
  EdS geometry, i.e., $V_{\textsc{out}}=a^{3}\ell^{3}$.}
 \end{center}
\end{figure}

For sufficiently large $t_{+}$, the areal radius $R$ 
of the dust shell becomes much smaller than 
the cosmological horizon scale $H^{-1}$. 
In this case, the the motion of the 
dust shell is well described by the Newtonian approximation. 
Maeda and Sato showed that for sufficiently large $t_{+}$, 
$R$ behaves as\cite{ref:Maeda} 
\begin{equation}
 R(t_{+})\propto t_{+}{}^{(15+\sqrt{17})/24}\sim t_{+}{}^{0.797}
\sim t_{\rm s-}{}^{0.797}.
\end{equation}
Hence we find that $R/t_{\rm s-}\propto t_{\rm s-}{}^{-0.203}\rightarrow0$ 
for $t_{\rm s-}\rightarrow \infty$. Using this result and
eq.~(\ref{eq:V-tilde}), we find 
${\tilde V}_{\rm in}\longrightarrow 1$ for 
$t_{\rm s-}\rightarrow\infty$, and hence for $t_{+}\rightarrow\infty$, 
\begin{equation}
 V\longrightarrow a^{3}(t_{+})\ell^{3}.
\end{equation}
This equation means that the volume expansion rate approaches to
the background value asymptotically, i.e.,
\begin{equation}
 \dfrac{\dot{V}}{V}\longrightarrow 
3\dfrac{\dot{a}}{a}=3\dfrac{\dot{a}_{\textsc{b}}}{a_{\textsc{b}}}
\text{~~for~}t\rightarrow \infty.
\end{equation}
The effect of inhomogeneities on the volume expansion rate 
vanishes after the dust shell 
becomes much smaller than the horizon scale $H^{-1}$
(see fig.~\ref{fig:ratio.after}). 
However, as in the case of $k_{0}>0$, 
the volume itself is different from the background 
value $a_{\textsc{b}}^{3}\ell^{3}$. By eq.~(\ref{eq:a-def}), we 
find that the asymptotic value of $V$ is larger than 
the background value $a_{\textsc{b}}^{3}\ell^{3}$. 
However it should be noted that there is an upper limit on the
asymptotic value of $V$. 
Since $k$ can be arbitrarily small in this model, 
$\sqrt{1-\chi^{2}k}$ can be made arbitrarily larger than unity
except at $\chi=0$ and $\chi=\chi_{\rm sc}$. Hence we obtain
\begin{equation}
a_{\textsc{b}}(t_{+})
>a(t_{+})\left(1-\dfrac{4\pi}{3\ell^{3}}\chi_{\rm sc}^{3}\right)^{1/3}
>a(t_{+})\left(1-\dfrac{\pi}{6}\right)^{1/3}, 
\end{equation}
where the last inequality is obtained by setting 
$\chi_{\rm sc}=\ell/2$. Using this inequality, we obtain
\begin{equation}
a_{\textsc{b}}^{3}(t_{+})\ell^{3}
<a^{3}(t_{+})\ell^{3} < \left(1-\dfrac{\pi}{6}\right)^{-1}
a_{\textsc{b}}^{3}(t_{+})\ell^{3}
\sim 2.10\times a_{\textsc{b}}^{3}(t_{+})\ell^{3}.
\end{equation}
In contrast with the case of $k_{0}>0$, the volume itself 
can not be so different from the background value.

\section{SUMMARY AND DISCUSSION}
\label{sec:summary}

We have investigated an effect of inhomogeneities on the volume expansion 
in modified Swiss-Cheese universe model.
We considered two cases; the inhomogeneities collapse into 
black holes ($k_{0}>0$), while inhomogeneities 
expands faster than the background volume expansion ($k_{0}<0$). 
When inhomogeneities can be treated as perturbations of 
Einstein-de Sitter universe, the volume expansion is decelerated 
due to the second order contribution of the perturbations 
in both models. This result agrees with Nambu's second order
perturbation analysis. 

Although the choice of background homogeneous universe is
straightforward in the case of $|k|\chi^{2}\ll 1$,
it is not in the case of non-linear situation.
We introduced the background homogeneous universe 
in order to satisfy the conditions;
the cubic region in the background homogeneous universe 
with the same rest mass have the same evolution of the volume 
as that of corresponding region in the modified Swiss-Cheese universe
in the limit of $t\rightarrow 0$.

In the case of non-linear fluctuation with $k_{0}>0$, we find that 
the volume expansion rate approaches to that
of background universe asymptotically.
Since the modified Swiss-Cheese model is an exact solution 
of Einstein equation, 
we can obtain the precise behavior of the volume expansion.
We set it so that for $t\rightarrow 0$, 
the temporal evolution agrees with background one. 
Then we found that for $t\rightarrow\infty$, 
the temporal evolution agrees with that of outer EdS universe.
We can see these asymptotic behaviors analytically 
but we have to rely only on the numerical method to obtain the behavior
of the intermediate stages.
From the result of the numerical calculation (fig.~\ref{fig:positive}), 
we found that volume expansion is decelerated by the inhomogeneities.
This behavior coincides with a result obtained by Nambu.
We note that his result is based on the perturbation theory 
but our result is not. 
In our highly non-linear example, the volume expansion rate becomes 
negative at the intermediate stage.
This result may be the characteristics of the effect of 
non-linear fluctuations which can not be treated by the perturbation 
method.

In the case of $k_{0}<0$, the shell crossing singularity appears
in the inhomogeneous region.
In this article, we assumed that a spherical dust shell forms 
after the shell crossing singularity appears. 
We find that the inhomogeneities decelerate the volume expansion 
before it appears (fig.~\ref{fig:ratio.before}).
This is consistent with the previous results by Nambu but the 
inhomogeneities cannot be treated as the perturbation of homogeneous
universe.
After a spherical dust shell forms, the volume expansion rate approaches 
to that of the background universe asymptotically. 
Our dust shell universe model (fig.~\ref{fig:configuration})
is a crude approximation.
Therefore the  behavior of the volume expansion obtained 
(fig.~\ref{fig:ratio.after}),
especially at the early stage after the shell crossing,
contains the influences of this crudeness. 
But the asymptotic behavior ($t\rightarrow \infty$) might be
free from this approximation.

We fixed the form of $k(\chi)$.
Temporal evolution of the volume depends on the form of $k(\chi)$
but the asymptotic behavior may be independent of the form of $k(\chi)$;
\begin{align*}
 V&\longrightarrow a^{3}\ell^{3} \text{~~for~~} 
  t\rightarrow \infty.
\end{align*}

%
\vskip0.5cm
{\large{\bf Acknowledgements}}

We are grateful to colleagues in Department of Physics, 
Osaka City University for helpful discussions.


\begin{thebibliography}{99}
%
\bibitem{ref:Weinberg} S.~Weinberg, {\it Gravitation and Cosmology} 
(Wiley, New York, 1973).
%
\bibitem{ref:Smoot} G.~F.~Smoot {\it et al}., Astrophys.~J.~Lett. 
{\bf 396}, L1 (1992).
%
\bibitem{ref:KS84} H.~Kodama and M.~Sasaki, 
Prog.~Theor.~Phys.~Suppl. {\bf 78}, 1 (1984)
%
\bibitem{Schmidt:1998ys} B.~P.~Schmidt {\it et al.}, 
Astrophys.\ J.\ {\bf 507}, 46 (1998). 
%
\bibitem{Perlmutter:1999np} 
S.~Perlmutter {\it et al.}, Astrophys.\ J.\ {\bf 517}, 565 (1999). 
%
\bibitem{ref:futamase89}
 T.~Futamase, Mon. Not. R. astr. Soc. {\bf 237}, 187 (1989).
%
\bibitem{ref:futamase97}
 T.~Futamase, Phys. Rev. {\bf D53}, 681 (1997).
%
\bibitem{ref:Tomita}
K.~Tomita, Prog. Theor. Phys. {\bf 37}, 831 (1967).
%
\bibitem{ref:Russ}
 H.~Russ, M.H.~Soffle, M.~Kasai and G.~B{\" o}rner, Phys. Rev. 
{\bf D56}, 2044 (1997). 
%
\bibitem{ref:Mukhanov}
V.~F.~Mukhanov, L.~R.~W.~Abramo and R.~H.~Brandenberger,
Phys. Rev. Lett. {\bf 78}, 1624 (1997).
%
\bibitem{ref:Abramo}
L.~R.~W.~Abramo and R.~H.~Brandenberger,
Phys. Rev. {\bf D56}, 3248 (1997).
%
\bibitem{ref:Nambu}
Y.~Nambu, Phys.~Rev. {\bf D62}, 104010 (2000).
%
\bibitem{ref:Nambu-2}
Y.~Nambu, Phys.~Rev. {\bf D63}, 044013 (2001).
%
\bibitem{ref:Nambu-3}
Y.~Nambu, Phys.~Rev. {\bf D65}, 104013 (2002).
%
\bibitem{ref:Geshnizjani}
G.~Geshnizjani and R.~H.~Brandenberger,
arXiv:gr-qc/0204074.
%
\bibitem{ref:Hellaby-Lake}
C.~Hellaby and K.~Lake, Astrophys.~J.~{\bf 290}, 381(1985), 
%
\bibitem{ref:Israel}
  W. Israel, Nuovo Cimento {\bf 44B} (1966), 1; \\
-----, Nuovo Cimento {\bf 48B} (1967), 463; \\
-----, Phys. Rev. {\bf 153} (1967), 1388. 
%
\bibitem{ref:Maeda}
 K.~Maeda and H.~Sato,  Prog.~ Theor.~Phys.~{\bf 70}, 772 (1983).\\
 K.~Maeda and H.~Sato,  Prog.~ Theor.~Phys.~{\bf 70}, 1276 (1983). 
\end{thebibliography}
\end{document}